# New Collisions to Improve Pollard's Rho Method of Solving the Discrete Logarithm Problem on Elliptic Curves

**Ammar Ali Neamah[1,2]**
[1] **Department of Computer Science, Faculty of Computer Science and Mathematics, University of Kufa, Najaf, Iraq**
[2] **Department of Mathematics, University of Reading, Whiteknights, Reading, UK**

## Abstract

It is true that different approaches have been utilised to accelerate the computation of discrete logarithm problem on elliptic curves with Pollard's Rho method. However, trapping in cycles fruitless will be obtained by using the random walks with Pollard's Rho. An efficient alternative approach that is based on new collisions which are reliant on the values $a_i$ , $b_i$ to solve this problem is proposed. This may requires less iterations than Pollard's Rho original in reaching collision. Thus, the performance of Pollard's Rho method is more efficiently because the improved method not only reduces the number of mathematical operations but these collisions can also applied on previous improvements which reported in the literature.

## Introduction

The technology of public key cryptography has been proposed by Diffie and Hellman in 1976 (**Diffie and Hellman, 1976**) whereas the elliptic curve cryptosystems (ECCs) that is dependent on this technology were first proposed separately by Koblitz and Miller in 1985 (**Miller, 1986; Koblitz, 1987**). The security of these cryptographic systems is reliant on the hardness of solving the discrete logarithm problem on elliptic curves (ECDLP). These schemes will be broken easily if this problem can be resolved efficiently. Despite the fact that there are several attacking methods to resolve ECDLP, Pollard's Rho method (**Pollard, 1980**) not only is at present known as the fastest algorithm to resolve the discrete logarithm problem on elliptic curves, but its parallelized variant as well because its mathematical operations is less than other methods like Baby-Step Giant-Step **(Shanks, 1971).** This encourages researchers to utilise from automorphism of the group **(Duursma et al., 1990)**, random walk on certain equivalence classes **(Wiener and Zuccherato, 1999; Gallant et al., 2000)**, parallelization **(Oorschot and Wiener, 1999)**, iteration function **(Teske, 1998; 2001)**, negation map **(Wang and Zhang, 2012)** or cycle detection **(Brent, 1980; Cheon et al., 2012; Ezzouak et al., 2014)** to improve this attacking method. This paper will provide a new approach by using the theorem that proposed by Sadkhan and Neamah in 2010 **(Sadkhan and Neamah, 2011)** to improve Pollard's Rho method which use alternative collisions to resolve the ECDLP. After that, the developed method will be analysed by giving examples.

**Background**

In this section, the mathematical background of elliptic curve cryptography and the Pollard Rho **(Pollard, 1980)** method that uses an iteration function to make a succession of elements will be presented. This technique utilises cycle detection to find collision or a match. This matching will contribute to the solution of ECDLP.

**Elliptic Curve Cryptography (ECC)**

Elliptic Curves over the prime field $F_p$ can be defined by the equation $y^2 = x^3 + ax + b$, where $4a^3 + 27b^2 \neq 0$. Different curves can be generated by changing the values $a$ and b which belongs to $F_p$. The set of points (x, y) that satisfy this equation equipped with the addition operation forms a group, denoted by $\mathbf{E}\,(F_p)$. This operation can be defined over elliptic curves with a particular point $\mathbf{O}_\infty$ which is called the identity or the point at infinity. If $P = (x_1, y_1)$ and $Q = (x_2, y_2)$ belongs to $\mathbf{E}\,(F_p)$, $P \oplus Q = R = (x_3, y_3)$ can be added as follows:

**(1)** If $P = \mathbf{O}_\infty$ , $Q = \mathbf{O}_\infty$ or $P = \mathbf{O}_\infty$ and $Q = \mathbf{O}_\infty$ then R respectively equal Q, P or $\mathbf{O}_\infty$.

**(2)** If $Q = \ominus P = (x_1, -y_1)$ and $P \oplus (\ominus P) = \mathbf{O}_\infty$.

**(3)** Otherwise $P \oplus Q = (x_3, y_3)$ where

$$x_3 = \lambda^2 - x_1 - x_2, \quad y_3 = \lambda \cdot (x_1 - x_3) - y_1$$

And

**2**

$$\lambda = \begin{cases} \dfrac{3x_1{}^2 + a}{2y_1} & \text{if } P = Q \\[3mm] \dfrac{y_2 - y_1}{x_2 - x_1} & \text{if } P \neq Q \end{cases}$$

In ECC$_s$, obtaining the public key is made out multiplying the private key that is a random number **k,** belongs to the interval [1, n − 1] with a point **P** which has order **n**. The point P is called the generator of a cyclic group $\langle P \rangle$ = {$\boldsymbol{O}_\infty$, P, [2]P, . . . , [n − 1]P}, and its order can be defined as the smallest positive scalar n such that [n]P = $\boldsymbol{O}_\infty$ which symbolized by **ord** (P). The strength of the ECC$_s$ security is reliant on the ECDLP. This problem can be defined as follows: Let P has order **n**, which belongs to the points of an elliptic curve defined over the field $F_p$ and a point Q $\in \langle P \rangle$, then finding **k** $\in$ [1, n − 1] such that Q = [k]P = $\underbrace{P \oplus P \oplus \; \cdot \; \cdot \; \cdot \; \oplus P}_{k \; \text{times}}$ is called the discrete log of Q to the base P which is symbolized by k = $\log_P Q$. Because k can be inferred from Q if the ECDLP is easy, so the difficulty of the ECDLP plays a crucial role in the security of these cryptographic systems (**Hankerson** *et al.***, 2004 ; Chee and Park, 2005).**

**Theorem**

Let Q a point belongs to group points that generated by P where **ord** (P) = n, and R can be calculated as follows: [$a$]P $\oplus$ [$b$]Q where $a$, $b \in$ [1, n − 1] then the following equations can be obtained (**Sadkhan and Neamah, 2011**) **:**

1) If R = $\boldsymbol{O}_\infty$ and gcd ($b$, n) = 1 then $\log_P Q = \dfrac{n - a}{b}$ mod n.

2) If R = P and gcd ($b$, n) = 1then $\log_P Q = \dfrac{1 - a}{b}$ mod n.

3) If R = Q and gcd (1 − $b$, n) = 1 then $\log_P Q = \dfrac{a}{1 - b}$ mod n.

4) If R = $\ominus$ P and gcd ($b$, n) = 1 then $\log_P Q = \dfrac{-1 - a}{b}$ mod n.

5) If R = $\ominus$ Q and gcd (− 1 − $b$, n) = 1 then $\log_P Q = \dfrac{a}{-1 - b}$ mod n.

**Pollard's Rho Method**

The Pollard idea is to find k which is satisfying Q = [k]P by dividing the group of points on an elliptic curve into three disjoint sets $S_1$, $S_2$ and $S_3$ which have an almost equal size. Define the original iteration function on a point R as follows:

$$f(R) = \begin{cases} R \oplus P, & \text{if } R \in S_1, \\ [2]R, & \text{if } R \in S_2, \\ R \oplus Q, & \text{if } R \in S_3, \end{cases} \qquad (1)$$

Start with a point $R_0 = [a_0]P \oplus [b_0]Q$ where $a_0$, $b_0 \in [1, n-1]$ are randomly chosen and make a sequence $R_i$ by using this function until the collision occurs. The sequence $R_i$ can be expressed as $[a_i]P \oplus [b_i]Q$, where the numbers $a_i$, $b_i \in [1, n-1]$ are calculated as following:

$$a_{i+1} = \begin{cases} (a_i + 1) \bmod n & \text{if } R_i \in S_1, \\ 2a_i \bmod n & \text{if } R_i \in S_2, \\ a_i, & \text{if } R_i \in S_3, \end{cases} \quad \text{and} \quad b_{i+1} = \begin{cases} b_i & \text{if } R_i \in S_1, \\ 2b_i \bmod n & \text{if } R_i \in S_2, \\ (b_i + 1) \bmod n & \text{if } R_i \in S_3, \end{cases}$$

Since the number of points that lies in the curve that form cyclic group is a finite, this sequence will not only become periodic after applying this function but will start to repeat. Upon detection of a matching, this is $R_i = R_j \Rightarrow [a_i]P \oplus [b_i]Q = [a_j]P \oplus [b_j]Q$, if gcd $(b_j - b_i, n) = 1$, k can be obtained as follows (**Wiener and Zuccherato, 1999**) :

$$k = \log_P Q = (\frac{a_i - a_j}{b_j - b_i}) \bmod n$$

**Example 1**

Let $y^2 = x^3 + 130x + 565$ an elliptic curve is defined over $F_{719}$, and let P = (312, 90) point has order 233 on elliptic curve points. The discrete logarithm of point Q = (475, 662) to the base P can be computed as follows:

The points of this curve are { $\boldsymbol{O}_\infty$, (0, 224), . . . . . . . . , (718, 282), (718, 437)}, the partitions of these points as follows:

$S_1 = \{ \boldsymbol{O}_\infty, (0, 224), (0, 495), . . . . . , (238, 302), (238, 417)\}$

$S_2 = \{(241, 293), (241, 426), . . . . . , (475, 57), (475, 662)\}$,

$S_3 = \{(481, 169), (481, 550), . . . . . , (718, 282), (718, 437)\}$.

Let $a_0, b_0 \in [1, n-1]$ are selected as follows: $a_0 = 2$, $b_0 = 87$ then $R_0$ can computed as follows: $[a_0]P \oplus [b_0]Q = [2](312, 90) \oplus [87](475, 662) = $ **(312, 90)**

Table (1) shows the intermediate steps of Pollard's Rho method.

| $i$ | $R_i$ | $a_i$ | $b_i$ |
|---|---|---|---|
| 0 | (312, 90) = P | 2 | 87 |
| 1 | (567, 38) | 4 | 174 |
| 2 | (56, 560) | 4 | 175 |
| 3 | (366, 290) | 5 | 175 |
| 4 | (665, 336) | 10 | 117 |
| 5 | (508, 598) | 10 | 118 |
| 6 | (400, 173) | 10 | 119 |
| 7 | (63, 276) | 20 | 5 |
| 8 | (422, 356) | 21 | 5 |
| 9 | (548, 457) | 42 | 10 |
| 10 | (676, 651) | 42 | 11 |
| 11 | (671, 150) | 42 | 12 |
| 12 | ( 312, 629) = ⊝ P | 42 | 13 |
| 13 | (567, 681) | 84 | 26 |
| 14 | (670, 259) | 84 | 27 |
| 15 | (636, 198) | 84 | 28 |
| 16 | (627, 111) | 84 | 29 |
| 17 | (630, 529) | 84 | 30 |
| 18 | (665, 336) | 84 | 31 |

Since $R_4 = R_{18} =$ (665, 336), then $[a_4]P \oplus [b_4]Q = [a_{18}]P \oplus [b_{18}]Q$

[10](312, 90) ⊕ [117](475, 662) = [84](312, 90) ⊕ [31](475, 90)

Which gives

$$k = \frac{a_4 - a_{18}}{b_{18} - b_4} \mod 233 = \frac{10 - 84}{31 - 117} \mod 233 = (159) \cdot 147^{-1} \mod 233 = 158$$

Thus $k = \log_P Q = 158$.

**Example 2**

Let $y^2 = x^3 + 250 x + 844$ an elliptic curve is defined over $F_{1009}$, and let $P = (909, 601)$ point has order 1007 on elliptic curve points. The discrete logarithm of point $Q = (134, 52)$ to the base P can be computed as follows:

The points of this curve are { $O_\infty$ , (2, 315), . . . . . . . . ., (1007, 423), (1007, 586)}, the partitions of these points as follows:

$S_1 =$ { $O_\infty$, (2, 315), (2, 694), . . . . . ., (335, 138), (335, 871)}
$S_2 =$ {(338, 328), (338, 681), . . . . . ., (688, 407), (688, 602)},
$S_3 =$ {(691, 67), (691, 942), . . . . . ., (1007, 423), (1007, 586)}.

Let $a_0, b_0 \in [1, n-1]$ are chosen as follows: $a_0 = 46$, $b_0 = 229$ then $R_0$ can calculated as follows: $[a_0]P \oplus [b_0]Q = [46](909, 601) \oplus [229](134 , 52) =$ **(981, 997)**

Table (2) shows the intermediate steps of Pollard's Rho method.

| i | $R_i$ | $a_i$ | $b_i$ |
|---|---|---|---|
| 0 | (981, 997) | 46 | 229 |
| 1 | **(909, 601) = P** | 46 | 230 |
| 2 | (376, 671) | 46 | 231 |
| 3 | (113, 377) | 92 | 462 |
| 4 | (387, 58) | 93 | 462 |
| 5 | (482, 151) | 186 | 924 |
| 6 | (990, 873) | 372 | 841 |
| 7 | (30, 821) | 372 | 842 |
| 8 | (449, 876) | 373 | 842 |
| 9 | (434, 695) | 746 | 677 |
| 10 | (585, 82) | 485 | 347 |
| 11 | (974, 474) | 970 | 694 |
| 12 | (621, 167) | 970 | 695 |
| 13 | (502, 420) | 933 | 383 |
| 14 | (841, 440) | 859 | 766 |
| 15 | (160, 95) | 859 | 767 |
| 16 | (696, 855) | 860 | 767 |
| 17 | (368, 23) | 860 | 768 |
| 18 | (890, 335) | 713 | 529 |
| 19 | (657, 202) | 713 | 530 |
| 20 | (776, 151) | 419 | 53 |
| 21 | (445, 612) | 419 | 54 |
| 22 | (164, 541) | 838 | 108 |
| 23 | (993, 794) | 839 | 108 |
| 24 | (382, 677) | 839 | 109 |
| 25 | (454, 843) | 671 | 218 |
| 26 | (128, 749) | 335 | 436 |
| 27 | (819, 254) | 336 | 436 |
| 28 | (168, 891) | 336 | 437 |
| 29 | (932, 481) | 337 | 437 |
| 30 | (818, 301) | 337 | 438 |
| 31 | (270, 996) | 337 | 439 |
| 32 | (809, 842) | 338 | 439 |
| 33 | (206, 940) | 338 | 440 |
| 34 | (15, 152) | 339 | 440 |
| 35 | (597, 348) | 340 | 440 |
| 36 | (14,1004) | 680 | 880 |
| 37 | (595, 332) | 681 | 880 |
| 38 | (449, 133) | 355 | 753 |
| 39 | (434, 314) | 710 | 499 |
| 40 | (585, 927) | 413 | 998 |
| 41 | (974, 535) | 826 | 989 |
| 42 | (607, 458) | 826 | 990 |
| 43 | (660, 171) | 645 | 973 |
| 44 | (72, 278) | 283 | 939 |
| 45 | (351, 287) | 284 | 939 |
| 46 | (662, 783) | 568 | 871 |
| 47 | (697, 56) | 129 | 725 |
| 48 | (164, 541) | 129 | 736 |

Since $R_{22} = R_{48} = $ **(164, 541)**, then $[a_{22}]P \oplus [b_{22}]Q = [a_{48}]P \oplus [b_{48}]Q$

$[838](909, 601) \oplus [108](134, 52) = [129](909, 601) \oplus [736](134, 52)$

Which gives

$$k = \frac{a_{22} - a_{48}}{b_{48} - b_{22}} \bmod 1007 = \frac{838 - 129}{736 - 108} \bmod 1007 = (709) \cdot 628^{-1} \bmod 1007 = 766$$

$$\text{Thus } k = \log_P Q = 766.$$

## Improving Pollard's Rho Method by Using New Equations

The main idea of alternative collision of pollard's Rho on elliptic curve group method is that looks for a new collision in the group. Using the equations has been proposed by Sadkhan and Neamah (**Sadkhan and Neamah, 2011**) in Theorem above can not only significantly contribute to improving this method by reducing the mathematical operations but collision-detection algorithms can also be applied. Elliptic curve points over field $F_p$ divide into three disjoint sets $S_1$, $S_2$ and $S_3$ with iterative function defined as follows:

$$f(R) = \begin{cases} R \oplus P, & \text{if } R \in S_1, \\ [2]\,R, & \text{if } R \in S_2, \\ R \oplus Q, & \text{if } R \in S_3, \end{cases} \tag{2}$$

The sequence $R_i$ can be expressed as $[a_i]P \oplus [b_i]Q$, where the numbers $a_i$, $b_i \in [1, \; n-1]$ are calculated as following:

$$a_{i+1} = \begin{cases} (a_i + 1) \bmod n & \text{if } R_i \in S_1, \\ 2a_i \bmod n & \text{if } R_i \in S_2, \\ a_i, & \text{if } R_i \in S_3, \end{cases} \quad \text{and} \quad b_{i+1} = \begin{cases} b_i & \text{if } R_i \in S_1, \\ 2b_i \bmod n & \text{if } R_i \in S_2, \\ (b_i + 1) \bmod n & \text{if } R_i \in S_3, \end{cases}$$

The collision-detection will occur with the following equations after applying the iteration function:

1) If $R_i = O_\infty$ and gcd $(b_i, n) = 1$    then $k = \log_P Q = \dfrac{n - a_i}{b_i} \bmod n$

2) If $R_i = P$ and gcd $(b_i, n) = 1$    then $k = \log_P Q = \dfrac{1 - a_i}{b_i} \bmod n.$

3) If $R_i = Q$ and gcd $(1 - b_i, n) = 1$ then $k = \log_P Q = \dfrac{a_i}{1 - b_i} \bmod n.$

4) If $R_i = \ominus P$ and gcd $(b_i, n) = 1$    then $k = \log_P Q = \dfrac{-1 - a_i}{b_i} \bmod n.$

5) If $R_i = \ominus Q$ and gcd $(-1 - b_i, n) = 1$ then $k = \log_P Q = \dfrac{a_i}{-1 - b_i} \bmod n.$

If such collisions cannot be obtained, the original collision $R_i = R_j$, or the reverse collision $R_i = \ominus R_j$ respectively can be applied as follows:

**i)** $\log_P Q$ can be obtained as follows: $\mathbf{k} = (\dfrac{a_i - a_j}{b_j - b_i}) \bmod n$, if gcd $(b_j - b_i, n) = 1$.

**ii)** $\log_P Q$ can be obtained as follows: $\mathbf{k} = (\dfrac{a_i + a_j}{-b_i - b_j}) \bmod n$, if gcd $(-b_i - b_j, n) = 1$.

## Example 3

From example (1) since $R_i = R_0 = P$ and gcd $(87, 233) = 1$ then condition (2) can be applied $k = \log_P Q = \dfrac{1 - a_0}{b_0} \bmod n = \dfrac{232}{87} \bmod 233 = (232) \cdot 87^{-1} \bmod 233 = 158.$

$$\text{Thus } k = \log_P Q = 158.$$

Or,

Since $R_i = R_{12} = \ominus P$ and gcd $(13, 233) = 1$ then the condition (4) can also be applied $k = \log_P Q = \dfrac{-1 - a_{12}}{b_{12}} \bmod n = \dfrac{190}{13} \bmod 233 = (190) \cdot 13^{-1} \bmod 233 = 158.$

$$\text{Thus } k = \log_P Q = 158.$$

## Example 4

From example (2) since $R_8 = \ominus R_{38}$ and gcd $(419, 1007) = 1$ then the condition (ii) can be applied $k = \dfrac{a_8 + a_{38}}{-b_8 - b_{38}} \bmod n = \dfrac{373 + 355}{-842 - 753} \bmod 1007 = (728) \cdot 419^{-1} \bmod 1007 = 766.$

$$\text{Thus } k = \log_P Q = 766.$$

Or,

Since $R_i = R_1 = P$ and gcd $(230, 1007) = 1$ then the condition (2) can also be applied $k = \log_P Q = \dfrac{1 - a_1}{b_1} \bmod 1007 = \dfrac{962}{230} \bmod 1007 = (962) \cdot 230^{-1} \bmod 1007 = 766.$

$$\text{Thus } k = \log_P Q = 766.$$

**Comparison between Methods**

Analysing the methods play a crucial role in computer programming, so knowing the best methods is depended on this analysis because there are many methods available to a specific application. The performance of these methods such as Pollard's Rho can be evaluated by computing mathematical operations. If such methods have fewer steps than others, running time will be the best. Therefore, the developed Pollard's Rho method that implemented in these examples have shown the mathematical operations can be reduced from 18 to 1 or 12 operations with respect to the first example, and from 48 to 2 operations or 38 in the worst case by using the reverse collision $R_i = \ominus R_j$ with respect to the second example. These examples have utilised prime numbers with size between three and four digits. The **tables 1 & 2** of these examples show that the intermediate steps of Pollard's Rho method. The experimental results for this method with its new equations have been certainly improved because the point Q belong to a cyclic group which generated by P will always contain the points P, Q, $\ominus$ P, $\ominus$ Q and $\boldsymbol{O_\infty}$ in this group. This implies these collisions not only have always occur but the probability of a collision has also increased five times compared with original method. It can be concluded that the proposed improved method is not only better than the original pollard's Rho method but these alternative collisions can also be applied to previous proposed improvements such that dividing the iteration function into about 20 sets **(Teske, 1998; 2001)**.

**Conclusion**

The improved Pollard's Rho method has been outlined by using new collisions. These collisions are reliant on values $a_0$ and $b_0$ that have been playing a significant role in reducing the total number of mathematical operations if these values are chosen carefully. This means selecting these numbers will significantly contribute to solve the ECDLP quickly. This modified method can be considered as an important addition to methods of improving the Pollard's Rho. Although these results are related to small prime finite fields, this can represent a good reduction of complexity when it is applied by using large prime numbers. The parallel collision search can also be applied by selecting different start points which have different values $a_0$ and $b_0$ for each processor in order to produce their own sequences of points, so the chance of collision will be considerably increased. This means the collision may not only occur with the points P, Q, $\ominus$ P , $\ominus$ Q and $\boldsymbol{O}_\infty$ more than once but also with the points $R_j$ and $\ominus R_j$ in the worst case.

## Acknowledgment

The author is grateful to Mr. Sebastian Watkins, who revised this article linguistically for his helpful suggestions that improved it.

## Ethics

This article is original and contains unpublished materials. The corresponding author confirms that all of the other authors have read and approved the manuscript and no ethical issues involved.